\documentclass[12pt]{article}
\usepackage{graphicx}
\usepackage{times}

\usepackage{amsmath}

\usepackage{epsf}
\def\beq{\begin{equation}}
\def\eeq{\end{equation}}
\def\bea{\begin{eqnarray}}
\def\eea{\end{eqnarray}}

\def\vel{\left|}
\def\ver{\right|}
\def\nnb{\nonumber}
\def\ga{\left(}
\def\dr{\right)}

\usepackage{slashed}
\def\rar{\rightarrow}
\def\nnb{\nonumber}

\def\ba{\begin{array}}
\def\ea{\end{array}}
\def\bea{\begin{eqnarray}}
\def\eea{\end{eqnarray}}

\def\ds{\displaystyle}

\def\vel{\left|}
\def\ver{\right|}
\def\nnb{\nonumber}
\def\ga{\left(}
\def\dr{\right)}

\def\rar{\rightarrow}
\def\nnb{\nonumber}

\def\lla{\left<}
\def\rra{\right>}

\def\es{\!\!\! &=& \!\!\!}

\def\ar{&+& \!\!\!}
\def\ek{&-& \!\!\!}
\begin{document}

\title{ {\Large {\bf Lepton Asymmetries for the $B_s \rar \gamma l^+ l^- $ Decay in a Family Non-Universal $Z'$ Model }}}
\author{ {\small B. B.  \c{S}irvanl{\i}}\\
{\small Gazi University, Faculty of Arts and Science, Department of Physics} \\
{\small 06100, Teknikokullar Ankara, Turkey}}

\thispagestyle{empty}
\maketitle

\begin{abstract}
The exclusive $B_s \rar \gamma l^+ l^- $ decay is analyzed in the
framework of a family non-universal Z'model by calculating the
differential branching ratio, double lepton polarizations and
forward-backward asymmetries. Our results are compared against
those of the Standard Model. The predictions of this work are
hoped to can be tested in the near future at LHCb.
\end{abstract}
~~~PACS number(s):12.60.--i, 13.20.--v, 13.20.He

%
\clearpage

\section{Introduction \label{s1}}

The rare decays in the SM proceed via the flavor changing neutral
current (FCNC) which are forbidden at the tree level. The rare
decays are one of the best grounds for testing the predictions of
the Standard Model (SM) at quantum level. Moreover, these decays
are also very promising for establishing new physics beyond the SM
indirectly. With operation of the LHCb, new windows are opened for
searching of rare decays. Recently, LHCb and CMS Collaborations
\cite{R01,R02} have announced the observation of $B_s \to \mu^+
\mu^-$ decays. This decay is helicity suppressed and its matrix
element is proportional to the lepton mass. The branching ratio
for the $\mu^+ \mu^-$ channel is $1.8\times 10^{-9}$ in the SM. In
this sense, the observation of this decay is a great achievement
in particle physics.

Another rare decay which can be measured in LHCb is the $B_s \to
\ell^+ \ell^- \gamma$ transition. The main feature of this decay
is that the helicity suppression is overcome. For this reason,
despite that the width of this decay has an extra factor of fine
structure constant $\alpha$, it is comparable to the decay width
of the pure leptonic $B_s \to \ell^+ \ell^-$ channel. Indeed, it
is shown in \cite{R03} that the $B_s \to \ell^+ \ell^-\gamma$
decay can have larger branching ratio compared to that of the $B_s
\to \ell^+ \ell^-$ channel. As it has already been noted, $B_s \to
\ell^+ \ell^-$, and consequently $B_s \to \ell^+ \ell^-\gamma$
decays are both sensitive to the existence of new physics beyond
the SM. One possible extension of the SM is the family
non-universal $Z^\prime$ model, which contains family
non-universal $U(1)$ gauge symmetries. Such type models appear in
some low energy manifestations of the string theory \cite{R04} and
$E_6$ models \cite{R05}. Detailed information about this model can
be found in \cite{R06}.

In the frame work of this model $B_q$-$\bar{B}_q$ mixing, $B \to
X_s \mu^+ \mu^-$, $B_s \to \mu^+ \mu^-$ decays \cite{R07}; $B \to
K^\ast \ell^+ \ell^-$ \cite{R08}, $B_s \to \phi \mu^+ \mu^-$
\cite{R09}, $B \to K_1 \ell^+ \ell^-$ \cite{R10}, $\Lambda_b \to
\Lambda \ell^+ \ell^-$ \cite{R11}, and $\Sigma_b \to \Sigma \mu^+
\mu^-$ \cite{R12} processes have already been investigated,
respectively. In the present work, we study the rare $B_s \to
\ell^+ \ell^- \gamma$ decay.

The work is organized as follows: In section 2, we present the
matrix element for the $B_s \to \ell^+ \ell^- \gamma$ decay. In
section 3 the expressions of the differential branching ratio,
double lepton polarization, as well as forward-backward
asymmetries are presented. The last section is devoted to the
numerical analysis and discussions.

\section{The matrix element for the $B_s \to \ell^+ \ell^- \gamma$ decay}

As it is well known, $B_s \to \ell^+ \ell^-$ decay is described by
the $b \to s \ell^+ \ell^-$  transition at the quark level. In the
SM, the effective Hamiltonian for the $b \to s \ell^+ \ell^-$
transition can be written in the following form \cite{R13,R14}:
\bea \label{e01} {\cal H}_{eff} \es {\alpha_{em} G_F \over 2
\sqrt{2} \pi} V_{tb} V_{ts}^\ast \Bigg\{ C_9^{eff}(\mu)
\Big[\bar{s} \gamma_\mu (1-\gamma_5) b \Big] \bar{\ell} \gamma_\mu
\ell + C_{10} (\mu) \Big[\bar{s} \gamma_\mu (1-\gamma_5) b\Big]
\bar{\ell} \gamma_\mu \gamma_5 \ell \nnb \\
\ek  2 C_7(\mu) {i\over q^2} m_b \Big[\bar{s} \sigma_{\mu\nu}
(1+\gamma_5) b\Big] \bar{\ell} \gamma_\mu \ell \Bigg\}~, \eea
where $V_{tb}$ and $V_{ts}^\ast$ are the elements of the
Cabibbo-Kobayashi-Maskawa (CKM) mixing matrix, $C_7(\mu)$,
$C_9^{eff}(\mu)$ and $C_9^{eff}(\mu)$ are the Wilson coefficients.
If the mixing between $Z$ and $Z^\prime$ is neglected, the
contribution coming from $Z^\prime$ can be described by just
modifying of the Wilson coefficients without introducing any new
operator structure. The expression of the effective Hamiltonian
which describe the contribution of the Z' boson. It can be written
in the following form \cite{R15,R16}: \bea \label{e02} {\cal
H}^{Z^\prime} \es - {2 G_F \over \sqrt{2}} V_{tb} V_{ts}^\ast
\Bigg[ {B_{sb}^L B_{\ell\ell}^L \over V_{tb} V_{ts}^\ast} \bar{s}
\gamma_\mu (1-\gamma_5) b \bar{\ell} \gamma_\mu
(1-\gamma_5) \ell \nnb \\
\ar {B_{sb}^L B_{\ell\ell}^R \over V_{tb} V_{ts}^\ast} \bar{s}
\gamma_\mu (1-\gamma_5) b \bar{\ell} \gamma_\mu (1+\gamma_5) \ell
\Bigg]~, \eea where $B_{sb}^L = \vel B_{sb}^L \ver
e^{i\varphi_S^L}$ and $B_{\ell\ell}^{L,R}$ correspond to the
interaction vertex of $Z^\prime$ with quark and leptons,
respectively.

In order to take into account the contributions coming from the
$Z^\prime$ boson it is sufficient to modify the Wilson
coefficients $C_9^{eff}(M_W)$ and $C_{10}(M_W)$ in Eq(1) and Eq(2)
as follows: \bea \label{e03} C_9^{eff} \to C_9^{tot} \es C_9^{eff}
- {4 \pi \over \alpha_S} (28.82) {B_{sb}^L \over V_{tb}
V_{ts}^\ast} (B_{\ell\ell}^L +
B_{\ell\ell}^R) \nnb \\
C_{10} \to C_{10}^{tot} \es C_{10} + {4 \pi \over \alpha_S}
(28.82) {B_{sb}^L \over V_{tb} V_{ts}^\ast} (B_{\ell\ell}^L -
B_{\ell\ell}^R)~, \eea where $\alpha_S$ is the strong coupling
constant. It should be noted here that $C_7$ receives no
contribution from $Z^\prime$ and evolution of $C_9^{tot}$ and
$C_{10}^{tot}$ from weak to $\mu=m_b$ scale should be the same as
in SM.

The Wilson coefficient $C_7^{eff}$ in the SM is given by
\cite{R17}:
  \bea
\label{e04} C_7^{eff}(m_b) \es \eta^{\frac{16}{23}}
C_7(\mu_W)+ \frac{8}{3} \left( \eta^{\frac{14}{23}}
-\eta^{\frac{16}{23}} \right) C_8(\mu_W)+C_2 (\mu_W) \sum_{i=1}^8
h_i \eta^{a_i}~, \eea where \bea
 C_2(\mu_W)=1~,~~ C_7(\mu_W)=-\frac{1}{2}
D_0(x_t)~,~~ C_8(\mu_W)=-\frac{1}{2} E_0(x_t)~ . \nnb \eea
$D_0(x_t)$ and $E_0(x_t)$ are to be functions of $(x_t)$ and $x_t
= m_{t}^{2}/m_W^{2}$. $m_{t}^{2}$ and $m_W^{2}$ are the top quark
and $W$ boson masses, respectively. $D_0(x_t)$ and $E_0(x_t)$ are
defined as:
 \bea \label{e05} D_0(x_t) \es - \frac{(8
x_t^3+5 x_t^2-7 x_t)}{12 (1-x_t)^3}
+ \frac{x_t^2(2-3 x_t)}{2(1-x_t)^4}\ln x_t~, \nnb \\ \nnb \\
\label{e0} E_0(x_t) \es - \frac{x_t(x_t^2-5 x_t-2)}{4 (1-x_t)^3} +
\frac{3 x_t^2}{2 (1-x_t)^4}\ln x_t~. \eea
The coefficients $a_i$ and $h_i$ in Eq. (\ref{e04})
are given as:
\bea \frac{}{}
\label{e06}
\begin{array}{rrrrrrrrrl}
a_i = (\!\! & \frac {14}{23}, & \frac {16}{23}, & \frac {6}{23}, &
- \frac {12}{23}, &
0.4086, & -0.4230, & -0.8994, & 0.1456 & \!\!)  \vspace{0.1cm}, \\
h_i = (\!\! & 2.2996, & - 1.0880, & - \frac{3}{7}, & -
\frac{1}{14}, & -0.6494, & -0.0380, & -0.0186, & -0.0057 & \!\!),
\end{array}
\eea
and the parameter $\eta$ is defined as:
 \bea \eta \es
\frac{\alpha_s(\mu_W)} {\alpha_s(\mu_b)}~,\nnb \eea
 with
 \bea \alpha_s(x)=\frac{0.118}{1-\ds\frac{23}{3}\ds\frac{
\alpha_s(m_Z)}{2\pi}\ln(\ds\frac{m_Z}{x})}.
\nnb \eea

The expression for the
Wilson coefficient $C_{9}^{eff}(\hat s)$ is given as \cite{R17}:
\bea \label{e07}
C_{9}^{eff}(\hat{s}) & = & C_9^{NDR}\eta(\hat s) + h(z, \hat
s)\left( 3 C_1 + C_2 + 3 C_3 + C_4 + 3
C_5 + C_6 \right) \nonumber \\
& & - \frac{1}{2} h(1, \hat s) \left( 4 C_3 + 4 C_4 + 3
C_5 + C_6 \right) \nonumber \\
& & - \frac{1}{2} h(0, \hat s) \left( C_3 + 3 C_4 \right) +
\frac{2}{9} \left( 3 C_3 + C_4 + 3 C_5 + C_6 \right), \eea

where $\hat{s}=q^2/m_b^2$ ,
$\hat{m_\ell}=m_\ell/m_B$, and
\bea
C_9^{NDR} & = & P_0^{NDR} +
\frac{Y_0(x_t)}{\sin^2\theta_W} -4 Z_0(x_t) + P_E E(x_t). \nnb \eea
Note that the small contribution coming from
$P_E$ is neglected in further numerical analysis. In the naive dimensional
regularization scheme we have $P_0^{NDR}=2.60 \pm 0.25$
\cite{R17}, and remaining two functions, $Y(x_t)$ and
$Z(x_t)$  are expressed as:

\bea
\label{e08}
Y_0(x_t) \es \frac{x_t}{8} \left[ \frac{x_t
-4}{x_t -1}+\frac{3 x_t}{(x_t-1)^2} \ln x_t \right]~, \nnb \\ \nnb \\
Z_0(x_t) \es \frac{18 x_t^4-163 x_t^3+259 x_t^2 -108 x_t}{144
(x_t-1)^3} +\left[\frac{32 x_t^4-38 x_t^3-15 x_t^2+18 x_t}{72
(x_t-1)^4} - \frac{1}{9}\right] \ln x_t~.
\eea The coefficients $ \eta(\hat s)$,
$\omega(\hat s)$, $h(y,\hat s)$, and $h(0, \hat s)$ in Eq. (\ref{e07})
are given as:
\bea \label{e09}
 \eta(\hat s) \es 1 + \frac{\alpha_{s}(\mu_{b}) \omega(\hat
s)}{\pi} \nnb \\ \nnb \\
 \omega(\hat s) \es - \frac{2}{9} \pi^2 -
\frac{4}{3}\mbox{Li}_2(\hat s) - \frac{2}{3}
(\ln \hat s) \ln(1-\hat s) - \frac{5+4\hat s}{3(1+2\hat s)}\ln(1-\hat s) - \nonumber \\
& &  \frac{2 \hat s (1+\hat s) (1-2\hat s)}{3(1-\hat s)^2 (1+2\hat
s)} \ln \hat s + \frac{5+9\hat s-6\hat s^2}{6 (1-\hat s) (1+2\hat
s)}~, \nnb \\ \nnb \\
h(y, \hat s) \es -\frac{8}{9}\ln\frac{m_b}{\mu_b} - \frac{8}{9}\ln y +
\frac{8}{27} + \frac{4}{9} x \nnb \\
& & - \frac{2}{9} (2+x) |1-x|^{1/2} \left\{
\begin{array}{ll}
\left( \ln\left| \frac{\sqrt{1-x} + 1}{\sqrt{1-x} - 1}\right| -
i\pi \right), &
\mbox{for } x \equiv \frac{4z^2}{\hat s} < 1 \nonumber \\
2 \arctan \frac{1}{\sqrt{x-1}}, & \mbox{for } x \equiv \frac
{4z^2}{\hat s} > 1,
\end{array}
\right. \nnb \\ \nnb \\
h(0, \hat s) \es \frac{8}{27} -\frac{8}{9} \ln\frac{m_b}{\mu_b} -
\frac{4}{9} \ln \hat s + \frac{4}{9} i\pi. \eea where   $y=1$ or
$z=m_c/m_b$ and $\mbox{Li}_2(\hat s) $ is the Spence function.

The remaining Wilson coefficients  $C_j~(j=1,...6) $ are given as:
 \bea \label{coeffs} C_j=\sum_{i=1}^8 k_{ji}
\eta^{a_i} \qquad (j=1,...6)~, \nnb \eea and the constants
$k_{ji}$
 have the values,
\bea\frac{}{}
\begin{array}{rrrrrrrrrl}
k_{1i} = (\!\! & 0, & 0, & \frac{1}{2}, & - \frac{1}{2}, &
0, & 0, & 0, & 0 & \!\!),  \vspace{0.1cm} \\
k_{2i} = (\!\! & 0, & 0, & \frac{1}{2}, &  \frac{1}{2}, &
0, & 0, & 0, & 0 & \!\!),  \vspace{0.1cm} \\
k_{3i} = (\!\! & 0, & 0, & - \frac{1}{14}, &  \frac{1}{6}, &
0.0510, & - 0.1403, & - 0.0113, & 0.0054 & \!\!),  \vspace{0.1cm} \\
k_{4i} = (\!\! & 0, & 0, & - \frac{1}{14}, &  - \frac{1}{6}, &
0.0984, & 0.1214, & 0.0156, & 0.0026 & \!\!),  \vspace{0.1cm} \\
k_{5i} = (\!\! & 0, & 0, & 0, &  0, &
- 0.0397, & 0.0117, & - 0.0025, & 0.0304 & \!\!) , \vspace{0.1cm} \\
k_{6i} = (\!\! & 0, & 0, & 0, &  0, &
0.0335, & 0.0239, & - 0.0462, & -0.0112 & \!\!).  \vspace{0.1cm} \\
\end{array} \nnb
\eea

The Wilson coefficient $C_9^{tot}$ receives also long distance
effect contribution coming from real $c\bar{c}$. In the
phenomenological Breit-Wigner ansatz, the long distance part
$Y_{LD}$ is defined as: \bea
\label{e10}
Y_{LD}=-\frac{3 \pi
C^{(0)}}{\alpha^{2}}\sum_{V_{i}=\psi(1s)...\psi(6s)}
\kappa_{i}\frac{\Gamma(V_{i}\rightarrow
\ell^+\ell^-)m_{V_{i}}}{q^2-m_{V_{i}}^{2}+i\Gamma_{V_{i}}m_{V_{i}}}\eea
where $C^{0}=3C_{1}+C_{2}+3C_{3}+C_{4}+3C_{5}+C_{6}$, and $\kappa_{i}$
is the phenomenological factor for the lowest two resonances which
is predicted to be $\kappa_{J/\psi}=2.3$\cite{R18}. After these preliminary
remarks we can now proceed to study on the problem under consideration.

As we already noted that the $B_s \to \ell^+ \ell^- \gamma$ decay
can be obtained from $B_s \to \ell^+ \ell^- $ which described by
$b \to s \ell^+ \ell^- $ transition, by radiating the photon from
any internal and external charged particles. Having the effective
Hamiltonian for the $b \to s \ell^+ \ell^- $ transition our next
problem is to find the matrix element for $B_s \to \ell^+ \ell^-
\gamma$ decay. We have following three different contributions:

a) the photon is emitted from the initial quark fields,

b) the photon is emitted from final charged leptons,

c) the photon is radiated from charged particles in the loop.

The contributions of the diagrams when photon is emitted from
internal charged particles is proportional to factor
$m_{l}^{2}/m_{W}^{2}$. For this reason this contribution can also
be safely neglected. The contributions of the diagrams when photon
is emitted from final state charged lepton (this part is so-called
Bremsstrahlung part is proportional to the lepton mass, which
follows from helicity arguments.)  The matrix element
corresponding to this contribution is

\bea \mathcal{M}_{\mathrm{1}}=\frac{\alpha G_F}{2\sqrt{2} \pi}
V_{tb }V_{ts}^* e \imath f_B C_{10} 2 m_l \Bigg[
\bar{l}\Bigg(\frac{\slashed{\varepsilon} \slashed{P}_B}{2p_1
k}-\frac{ \slashed{P}_B \slashed{\varepsilon}}{2p_2 k}\Bigg)
\gamma_5 l \Bigg]\eea where $P_B$ is the B meson momentum, $f_B$
is the decay constant of the B meson.

The matrix element for the $B_s \to \ell^+ \ell^- \gamma $ decay
when photon is radiated from initial quarks can be written as:
\bea \label{e11} \mathcal{M}_{\mathrm{2}}=\lla \gamma \vel {\cal
H}_{eff} \ver B \rra \es {\alpha G_F \over 2 \sqrt{2} \pi} V_{tb}
V_{ts}^\ast \Bigg\{ C_9^{tot} \bar{\ell} \gamma_\mu \ell \lla
\gamma(k) \vel \bar{s} \gamma_\mu
(1-\gamma_5) b \ver B(p+k) \rra \nnb \\
\ar C_{10}^{tot} \bar{\ell} \gamma_\mu \gamma_5 \ell \lla \gamma(k) \vel \bar{s}
\gamma_\mu (1-\gamma_5)b \ver B(p+k) \rra \nnb \\
i \ek 2 C_7 {m_b\over q^2} \bar{\ell} \gamma_\mu \ell \lla \gamma(k) \vel \bar{s}
\sigma_{\mu\nu} q_\nu(1+\gamma_5) b \ver B(p+k) \rra
\Bigg\}~.
\eea

 The matrix elements in the above-expression are
defined in terms of the form factors as follows: \bea \label{e12}
\lla \gamma(k) \vel \bar{s} \gamma_\mu (1-\gamma_5) b \ver B(p+k)
\rra \es {e \over m_B^2} \Big\{ \epsilon_{\mu\nu\lambda\sigma}
\varepsilon_\mu^\ast
q_\lambda k_\sigma g(q^2)\nnb \\
\ar i [ \varepsilon_\mu^\ast (k\!\cdot\! q) - (\varepsilon^\ast \!\cdot\! q) k_\mu ]
f(q^2)\Big\} ~, \nnb \\ \nnb \\
\lla \gamma(k) \vel \bar{s}
i \sigma_{\mu\nu} q_\nu (1+\gamma_5) b \ver B(p+k) \rra \es
{e \over m_B^2} \Bigg\{ \epsilon_{\mu\nu\lambda\sigma} \varepsilon_\nu^\ast
q_\lambda k_\sigma g_1(q^2) \nnb \\
\ar i [ \varepsilon_\mu^\ast (k\!\cdot\! q) - (\varepsilon^\ast
\!\cdot\! q) k_\mu ] f(q^2)]\Big\} ~, \eea

where $\varepsilon^\ast_\mu$ and $k_\mu$  are the four vector
polarization and the four vector momentum of the photon
respectively, and $g(q^2)$, $f(q^2)$, $g_1(q^2)$ and $f_1(q^2)$
are the transition form factors.

The matrix element for the $B_s \to \ell^+ \ell^- \gamma $ decay
is the sum of $\mathcal{M}_{\mathrm{1}}$ and
$\mathcal{M}_{\mathrm{2}}$, i.e.
$\mathcal{M}=\mathcal{M}_{\mathrm{1}}+\mathcal{M}_{\mathrm{2}}$.
Using equations (11), (12) and (13) for the differential decay
width we get

\bea && \frac{d \Gamma}{d \hat{s}}=\nnb
\\ && \vel \frac{\alpha G_F}{2 \sqrt{2} \, \pi} V_{tb} V_{ts}^*
\ver^2 \, \frac{\alpha}{\ga 2 \, \pi \dr^3}\, m_B^5 \pi  \Bigg{\{}
\frac{1}{12} \, \int_\delta^{1-4 r} x^3 \,dx \, \sqrt{1-\frac{4
r}{1 - x}} \, m_B^2 \Bigg[ \ga \vel A \ver^2 + \vel B \ver^2
\dr \ga 1- x+ 2 r \dr \nnb \\
&&+ \ga \vel C \ver^2 + \vel D \ver^2 \dr \ga 1- x - 4 r \dr
\Bigg] - 2 C_{10} f_B r \int_\delta^{1-4 r} x^2 \,dx \, {\rm Re}
\ga A \dr \, {\rm ln} \displaystyle{\frac{1 +
\sqrt{1-\displaystyle{\frac{4 r}{1 - x}}}} {1 -
\sqrt{1-\displaystyle{\frac{4 r}{1 - x}}}}}
\nnb \\
&&- 4 \vel f_B~ C_{10} \ver^2 r \, \frac{1}{m_B^2} \,
\int_\delta^{1-4 r} dx \Bigg[ \ga 2 + \frac{4 r }{x} -\frac{2}{x}
-x \dr\, {\rm ln} \displaystyle{\frac{1 +
\sqrt{1-\displaystyle{\frac{4 r}{1 - x}}}} {1 -
\sqrt{1-\displaystyle{\frac{4 r}{1 - x}}}}}
\nnb \\
&&+ \frac{2}{x} \ga 1-x \dr \, \sqrt{1-\frac{4 r}{1 - x}}\, \Bigg]
\Bigg{\}}~, \eea where the $f_B$ is the leptonic decay constant of
the $B$ meson, $x=\displaystyle{\frac{2 E_\gamma}{m_B}}$ is a
dimensionless parameter with $E_\gamma$ being the photon energy
and $r=\displaystyle{\frac{m_l^2}{m_B^2}}$. The lower limit of
integration over $x$ comes from imposing a cut $\delta$ on the
photon energy (for details see \cite{R03}).

  In further numerical analysis we
shall use the results of \cite{R03} for the form factors which are
given as:

\bea \label{13} g(q^2) = \frac{1GeV}{(1-\frac{q^2}{5.6^2})^2},
f(q^2) = \frac{0.8GeV}{(1-\frac{q^2}{6.5^2})^2}~,\nnb \\
g_{1}(q^2) = \frac{3.74 GeV^2}{(1-\frac{q^2}{40.5})^2}, f_{1}(q^2)
= \frac{0.68 GeV^2}{(1-\frac{q^2}{30})^2}~. \eea

\begin{figure}
\includegraphics[scale=0.5]{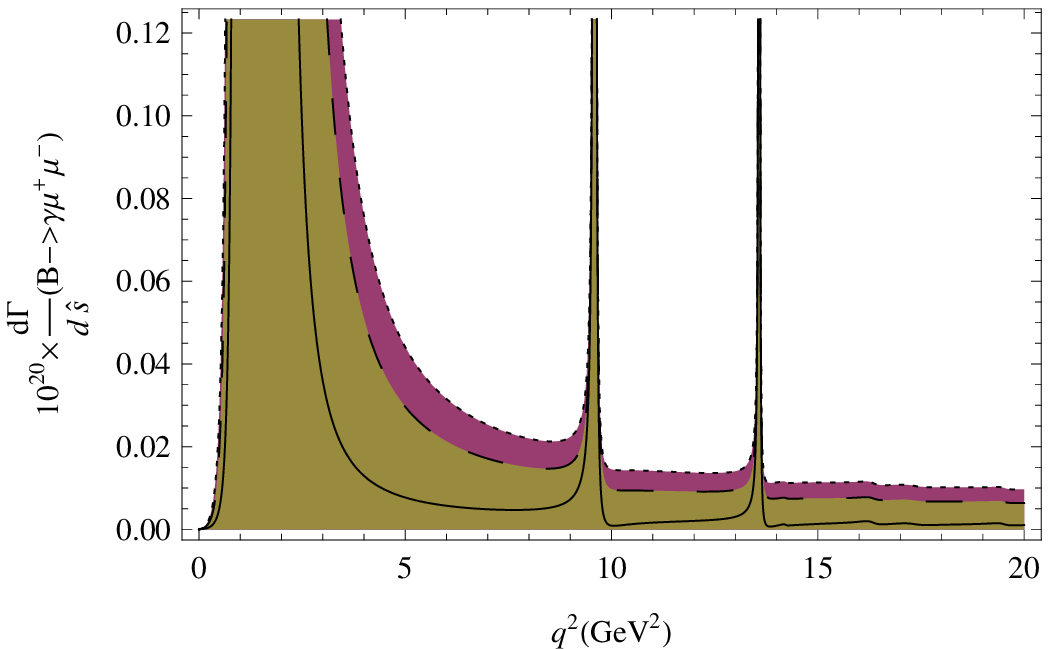}
\includegraphics[scale=0.5]{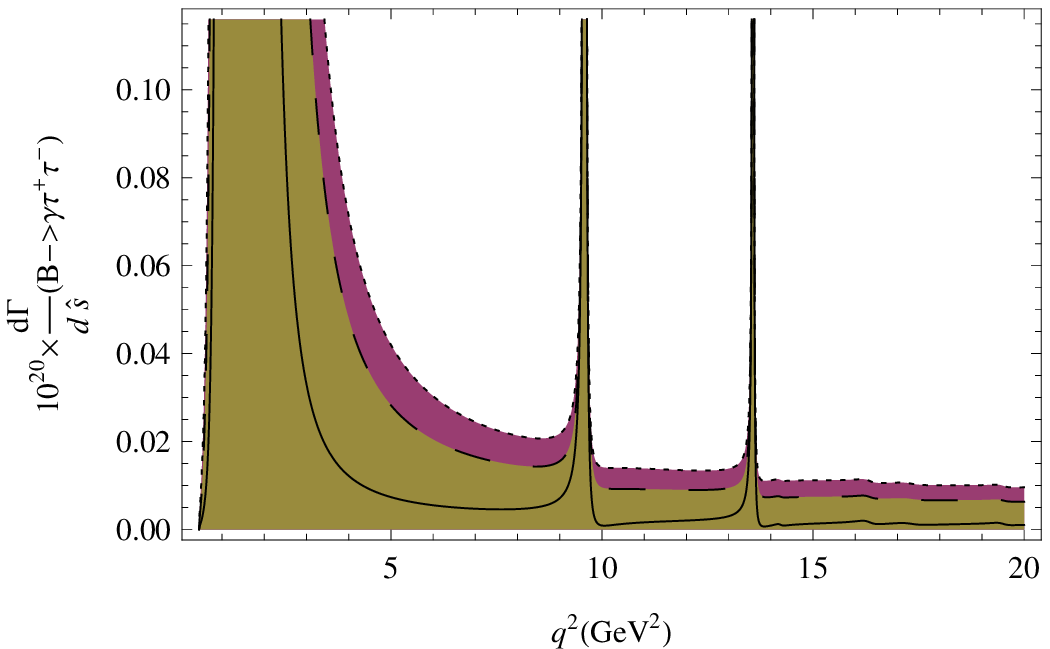}
\includegraphics[scale=0.5]{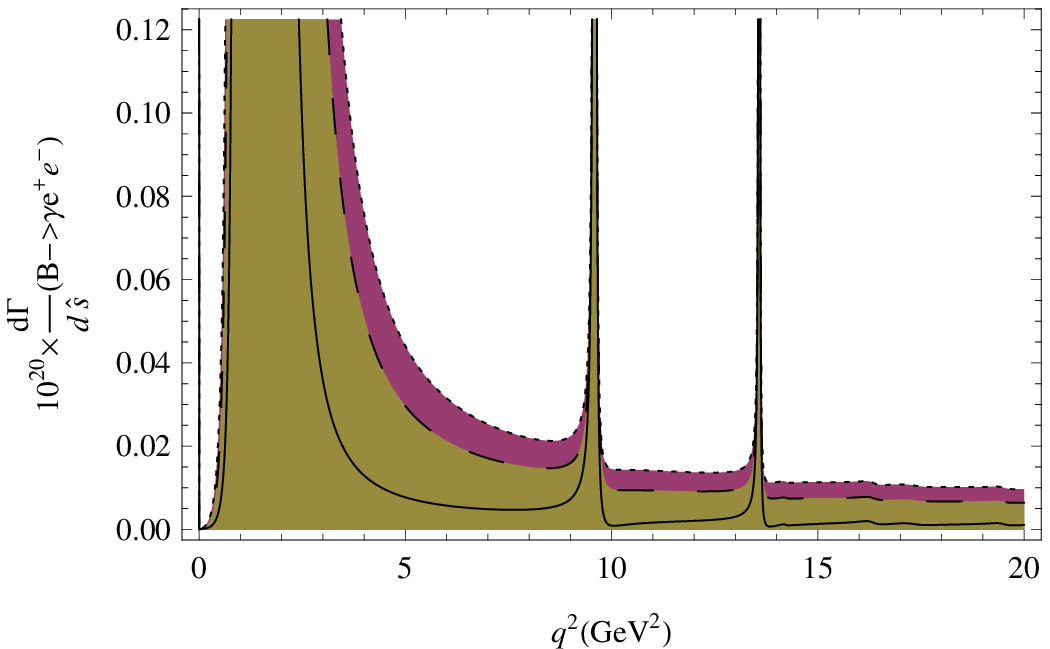}
\caption{ The dependence of the differential decay rate of the $B
\rar \gamma l^+ l^- $ on $q^2$ for $\mu$, $\tau$, e
leptons.\label{fig1} }
\end{figure}

Since $Z^\prime$ boson contributions are introduced by just modifying the
Wilson coefficients $C_9^{eff}$ and $C_{10}$, the expressions of the doubly
and singly polarized leptons are the same as given in \cite{R19,R20,R21},
whose  explicit expressions are as follows:

\bea \label{e15} P_{LL} (\hat{s})\es
\frac{1}{\Delta(\hat{s})} \nnb \\&\times&\Bigg\{ \frac{1}{2} f_B^2
m_B^4 \Big\{ (1-\hat{s})^2 ({\cal I}_1+{\cal I}_4) - [2 \hat{s} +
(1+\hat{s}^2) v^2] {\cal I}_3 + [2 \hat{s} - (1+\hat{s}^2) v^2]
{\cal I}_6 \Big\}
\vel F \ver^2 \nnb \\
&-& \frac{1}{2 \hat{m}_\ell} f_B m_B \hat{s} \Big[ 8 (1+\hat{s})
v^2 + m_B^2 (1-\hat{s}) (2-2 \hat{s} -2 v^2 + 2 \hat{s} v^2 +
v^4 +\hat{s} v^4) {\cal I}_{8} \nnb \\
\ek m_B^2 (1-\hat{s}^2) v^2 {\cal I}_{9}] \Big] \mbox{\rm Re}
[(A_1^\ast + B_1^\ast) F] - \frac{1}{3 \hat{m}_\ell^2} m_B^2
\hat{s}^2 (1-\hat{s})^2 (1-v^2)^2
\mbox{\rm Re} [A_1^\ast B_1 + A_2^\ast B_2] \nnb \\
\ek \frac{2}{3} m_B^2 \hat{s} (1-\hat{s})^2 (1+3 v^2) \Big( \vel
A_1 \ver^2 + \vel A_2 \ver^2 + \vel B_1 \ver^2 + \vel B_2 \ver^2
\Big) \Bigg\}~,
\eea
\bea
\label{e16} P_{LN} (\hat{s})\es
\frac{1}{\Delta(\hat{s})} \Bigg\{
 f_B m_B^3 \sqrt{\hat{s}} (1-\hat{s}^2) v^2
\mbox{\rm Im} [A_1^\ast F - B_1^\ast F] {\cal I}_7 \nnb \\ &-& 4
\pi f_B m_B \sqrt{\hat{s}} (1-\hat{s}) (1-\sqrt{1-v^2}) \mbox{\rm
Im} [(A_2^\ast + B_2^\ast) F] \Bigg\}~,
\eea
\bea
\label{e17} P_{NL} (\hat{s})\es \frac{1}{\Delta(\hat{s})}
\Bigg\{ f_B m_B^3 \sqrt{\hat{s}} (1-\hat{s}^2) v^2
\mbox{\rm Im} [-A_1^\ast F + B_1^\ast F] {\cal I}_7 \nnb \\
&+& 4 \pi f_B m_B \sqrt{\hat{s}} (1-\hat{s}) (1-\sqrt{1-v^2})
\mbox{\rm Im} [- (A_2^\ast + B_2^\ast) F] \Bigg\}~,
\eea
\bea
\label{e18} P_{LT}(\hat{s}) \es \frac{1}{\Delta(\hat{s})}
\Bigg\{ -\frac{1}{\sqrt{\hat{s}}} f_B^2 m_B^4 \hat{m}_\ell
(1-\hat{s}) v \Big[(1+\hat{s}) \vel F \ver^2 \Big]
({\cal I}_2+{\cal I}_4) \nnb \\
&+& \frac{4}{v} \pi f_B m_B \sqrt{\hat{s}} (1-\hat{s})
(1-\sqrt{1-v^2})
\mbox{\rm Re}[ (A_2^\ast - B_2^\ast) F] + 2 m_B \hat{m}_\ell \mbox{\rm Re}[A_1^\ast A_2 - B_1^\ast B_2] \Big] \nnb \\
&-& \frac{4}{v} \pi f_B m_B \sqrt{\hat{s}} (1+\hat{s})
(1-\sqrt{1-v^2}) \mbox{\rm Re}[(A_1^\ast + B_1^\ast) F] \Big]
\Bigg\}~,
\eea
\bea
\label{e19} P_{TL} (\hat{s}) \es
\frac{1}{\Delta(\hat{s})} \Bigg\{
- \frac{1}{\sqrt{\hat{s}}} f_B^2 m_B^4 \hat{m}_\ell (1-\hat{s}) v
\Big[(1+\hat{s}) \vel F \ver^2 \Big]
({\cal I}_2+{\cal I}_4) \nnb \\
&-&\frac{4}{v} \pi f_B m_B \sqrt{\hat{s}} (1-\hat{s})
(1-\sqrt{1-v^2}) \mbox{\rm Re}[ (A_2^\ast - B_2^\ast) F] - 2 m_B
\hat{m}_\ell \mbox{\rm Re}[A_1^\ast A_2 - B_1^\ast B_2] \Big] \nnb \\
&-&\frac{4}{v} \pi f_B m_B \sqrt{\hat{s}} (1+\hat{s})
(1-\sqrt{1-v^2})\mbox{\rm Re}[(A_1^\ast + B_1^\ast) F] \Big]
\Bigg\}~,
\eea
\bea
\label{e20} P_{NT}(\hat{s})  \es
\frac{1}{\Delta(\hat{s})} \Bigg\{ 2 f_B m_B^3 \hat{m}_\ell
(1-\hat{s})^2 v
\mbox{\rm Im}[ -A_1^\ast F + B_1^\ast F]({\cal I}_{8}-{\cal I}_{9}) \nnb \\
&-& 2 f_B m_B^3 \hat{m}_\ell (1-\hat{s}^2) v \mbox{\rm
Im}[(A_2^\ast+B_2^\ast) F] ({\cal I}_{8}-{\cal I}_{9}) \nnb
\\ &-& \frac{8}{3} m_B (1-\hat{s})^2 v \mbox{\rm Im}[- m_B \hat{s}
(A_1^\ast B_1 + A_2^\ast B_2)] \Bigg\}~,
\eea
\bea
\label{e21} P_{TN}(\hat{s})\es
\frac{1}{\Delta(\hat{s})} \Bigg\{ 2 f_B m_B^3 \hat{m}_\ell
(1-\hat{s})^2 v \mbox{\rm Im}[ A_1^\ast F - B_1^\ast F]
({\cal I}_{8}-{\cal I}_{9}) \nnb \\
&-& 2 f_B m_B^3 \hat{m}_\ell (1-\hat{s}^2) v
\mbox{\rm Im}[(A_2^\ast+B_2^\ast) F] ({\cal I}_{8}-{\cal I}_{9}) \nnb \\
&+&\frac{8}{3} m_B (1-\hat{s})^2 v \mbox{\rm Im}[- m_B \hat{s}
(A_1^\ast B_1 + A_2^\ast B_2)]\Bigg\}~,
\eea
\bea
\label{e22} P_{NN} (\hat{s}) \es
\frac{1}{\Delta(\hat{s})} \Bigg\{ f_B^2 m_B^4 \hat{s} \Big[
(1+v^2) {\cal I}_3 - (1-v^2) {\cal I}_6 \Big]
\vel F \ver^2 \nnb \\
&+& \frac{4}{3} m_B^2 \hat{s} (1-\hat{s})^2 v^2 \Big( 2 \mbox{\rm
Re}[A_1^\ast B_1 + A_2^\ast B_2 ] \Big)\Bigg\}~,
\eea
\bea
\label{e23} P_{TT}(\hat{s})  \es
\frac{1}{\Delta(\hat{s})} \Bigg\{ \frac{1}{2} f_B^2 m_B^4 \Big\{ -
(1-\hat{s})^2 (1-v^2) {\cal I}_1
+ [1 -v^2 - 4 \hat{s} + \hat{s}^2 (1-v^2)] {\cal I}_3 \nnb \\
\ek (1-v^2) (1-\hat{s})^2 {\cal I}_4 + (1-v^2) (1-\hat{s}^2) {\cal
I}_6 \Big\}
\vel F \ver^2 \nnb \\
\ek 4 f_B m_B^3 \hat{m}_\ell (1-\hat{s})^2
\mbox{\rm Re}[(A_1^\ast + B_1^\ast) F] ({\cal I}_{8}-{\cal I}_{9}) \nnb \\
&+& m_B \hat{m}_\ell \Big( \vel A_1 \ver^2 + \vel A_2 \ver^2 +\vel
B_1 \ver^2 +\vel B_1 \ver^2 \Big) \Big] \nnb \\
\ar \frac{8}{3} m_B^2
(1-\hat{s})^2 \Big( \hat{s} \mbox{\rm Re}[A_1^\ast B_1 + A_2^\ast
B_2] \Big) \Bigg\}~,
\eea
where, \bea \Delta (\hat{s}) \es
16 m_B \hat{m}_\ell (1-\hat{s})^2 \Big(\mbox{\rm Re}[m_B
\hat{m}_\ell (A_1^\ast B_1 + A_2^\ast B_1)] \Big) \nnb \\ &+&
\frac{2}{3} (1-\hat{s})^2 \Big[m_B^2 \hat{s} (3+v^2) \Big( \vel
A_1 \ver^2 + \vel A_2 \ver^2 + \vel B_1 \ver^2 + \vel B_2 \ver^2 \Big) \Big] \nnb \\
&-& \frac{1}{2} f_B^2 m_B^4 \vel F \ver^2 \Big\{ (1-\hat{s})^2 v^2
({\cal I}_1 + {\cal I}_4) - (1+\hat{s}^2 + 2 \hat{s} v^2) {\cal
I}_3 - [1-\hat{s} (4 - \hat{s} -2 v^2)] {\cal I}_6 \Big\} \nnb \\
&+& 2 f_B m_B \hat{m}_\ell \mbox{\rm Re} [(A_1^\ast + B_1^\ast) F]
\Big[ 8 (1+\hat{s}) + m_B^2 (1-\hat{s}^2) v^2 {\cal I}_{8} + m_B^2
(1-\hat{s}) (1-3\hat{s}) {\cal I}_{9} \Big] ~. \nnb \eea

In the expressions given above,
$v=\sqrt{1-4 \hat{m}_\ell^2/\hat{s}}$ is the lepton velocity, and

\bea \label{e6810}
A_1 = A_1 (\hat{s}) &=& \frac{-2 C_{7}^{eff}(\hat{s})}{q^2} \Big(m _{b} + m _{s} \Big) g_1(q^2) +  (C_{9}^{eff}(\hat{s})-C_{10}(\hat{s})) g(q^2) ~, \nnb \\
A_2 = A_2 (\hat{s}) &=& \frac{-2 C_{7}^{eff}(\hat{s})}{q^2} \Big(m _{b} - m _{s} \Big) f_1(q^2) +  (C_{9}^{eff}(\hat{s})-C_{10}(\hat{s})) f(q^2) ~, \nnb \\
B_1 = B_1 (\hat{s}) &=& \frac{-2 C_{7}^{eff}(\hat{s})}{q^2} \Big(m _{b} + m _{s} \Big) g_1(q^2) +  (C_{9}^{eff}(\hat{s})+C_{10}(\hat{s})) g(q^2) ~, \nnb \\
B_2 = B_2 (\hat{s}) &=& \frac{-2 C_{7}^{eff}(\hat{s})}{q^2}
\Big(m_{b}-m_{s} \Big) f_1(q^2) + (C_{9}^{eff}(\hat{s})+C_{10}(\hat{s})) f(q^2) ~, \nnb \\
 F =F(\hat{s})&=& 4 m_\ell C_{10}(\hat{s}), \nnb \eea
where ${\cal I}_i$ is determined as:
\bea {\cal I}_i = \int_{-1}^{+1} {\cal
F}_i(z) dz~,\nnb \eea where \bea
\begin{array}{lll}
{\cal F}_{1}  = \ds\frac{z^2}{(p_1 \cdot k) (p_2\cdot k)}~,& {\cal
F}_{2} = \ds\frac{z}{(p_1 \cdot k) (p_2 \cdot k)}~,&
{\cal F}_{3}  = \ds\frac{1}{(p_1 \cdot k) (p_2 \cdot k)}~, \nnb \\ \nnb \\
{\cal F}_{4}  = \ds\frac{z^2}{(p_1 \cdot k)^2}~,& {\cal F}_{5}  =
\ds\frac{z}{(p_1 \cdot k)^2}~,&
{\cal F}_{6}  = \ds\frac{1}{(p_1 \cdot k)^2}~, \\ \\
{\cal F}_{7}  = \ds\frac{z}{(p_2 \cdot k)^2}~,& {\cal F}_{8} =
\ds\frac{z^2}{p_1 \cdot k}~,& {\cal F}_{9} = \ds\frac{1}{p_1 \cdot
k}~.
\end{array} \nnb
\eea

\begin{figure}
\includegraphics[scale=0.8]{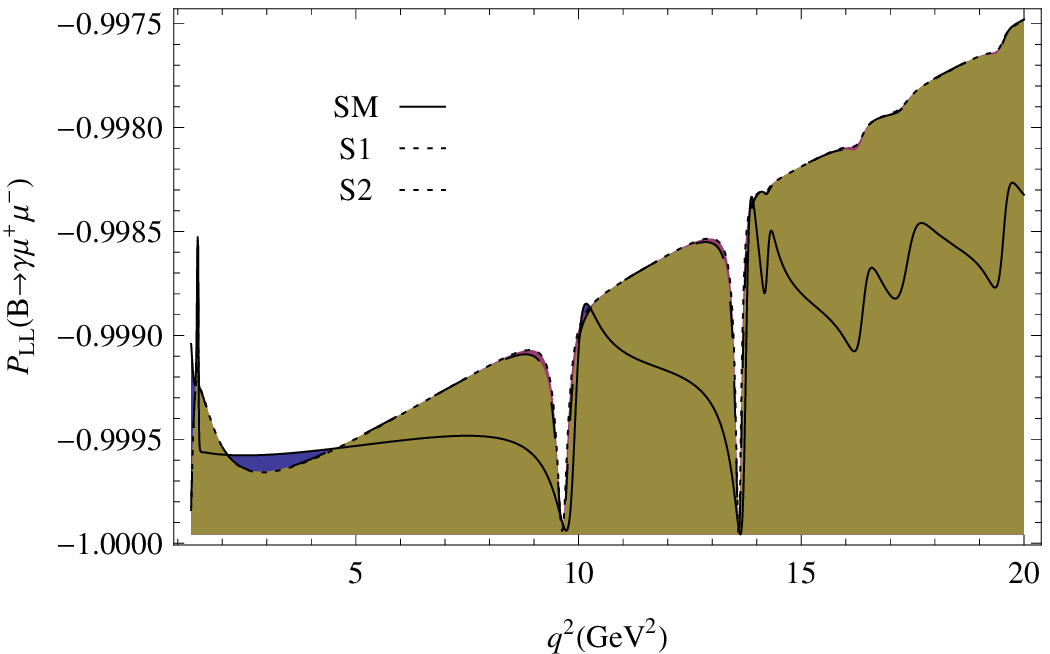}
\includegraphics[scale=0.8]{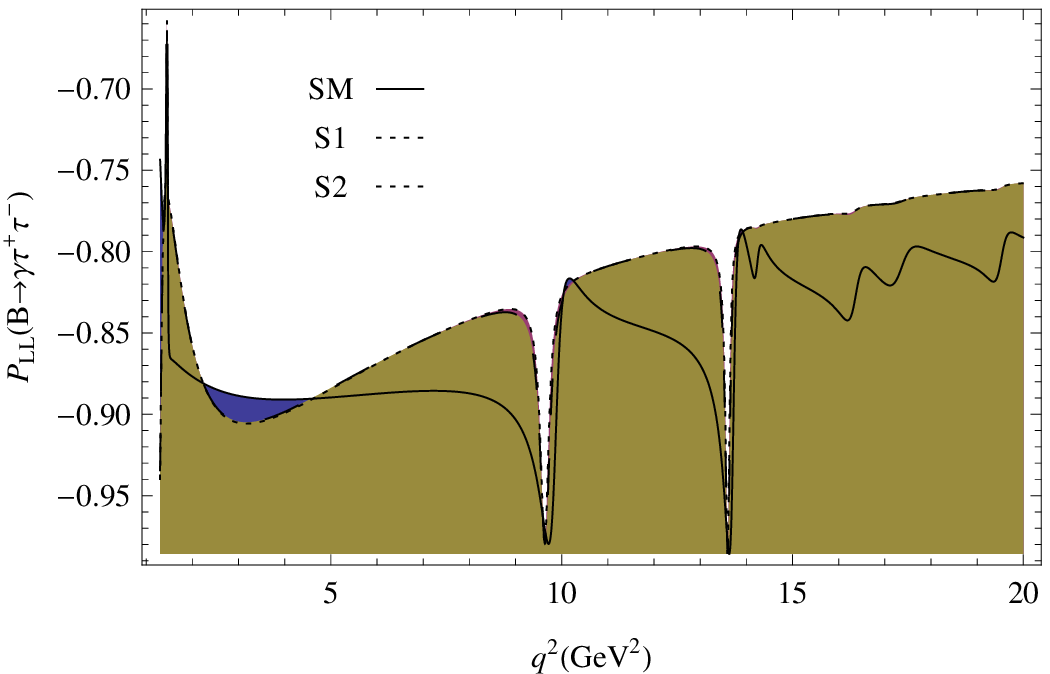}
\includegraphics[scale=0.8]{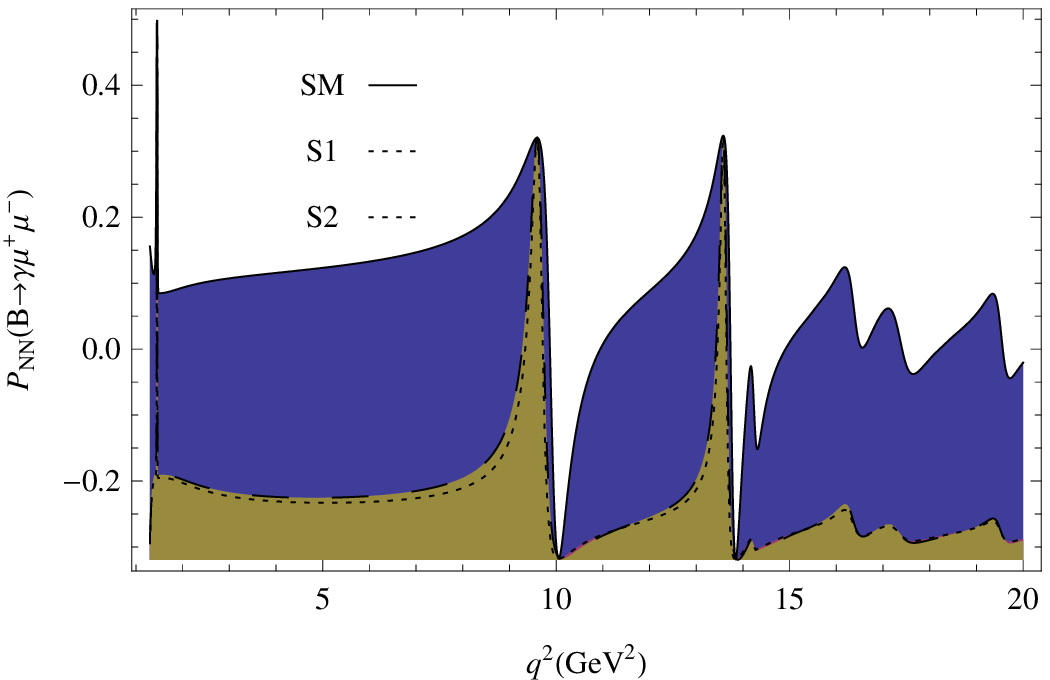}
\includegraphics[scale=0.8]{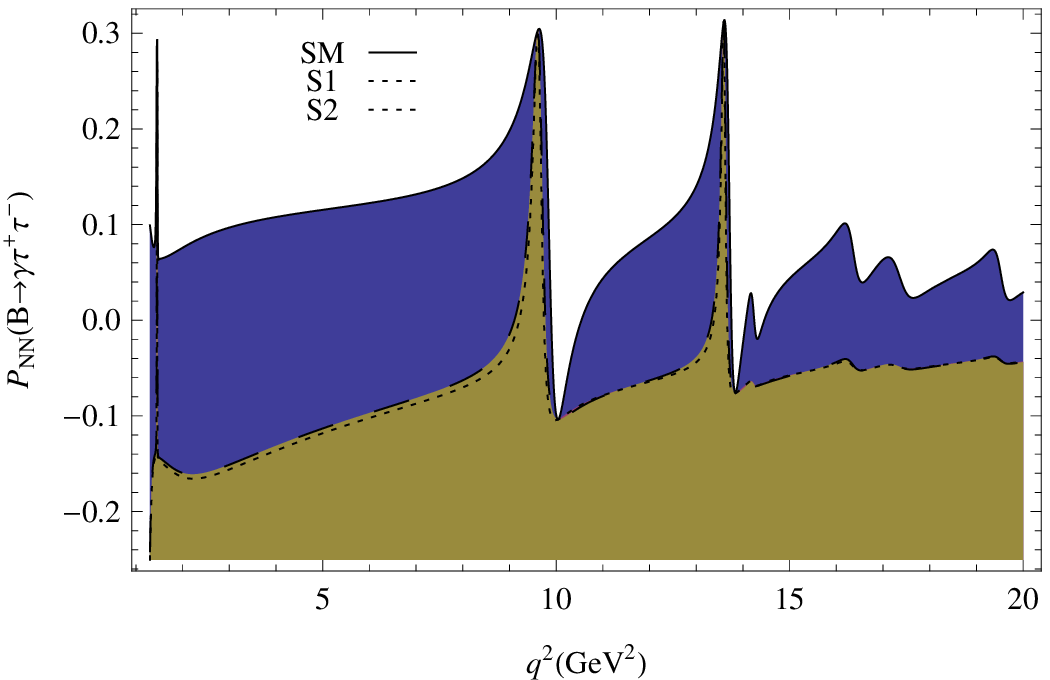}
\includegraphics[scale=0.8]{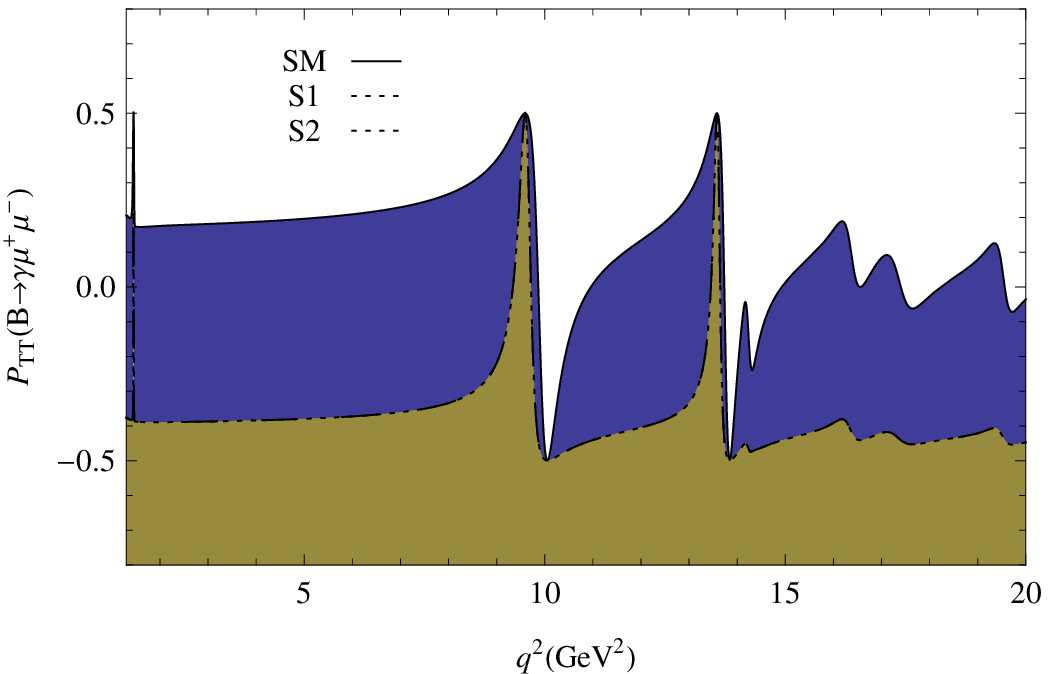}
\includegraphics[scale=0.8]{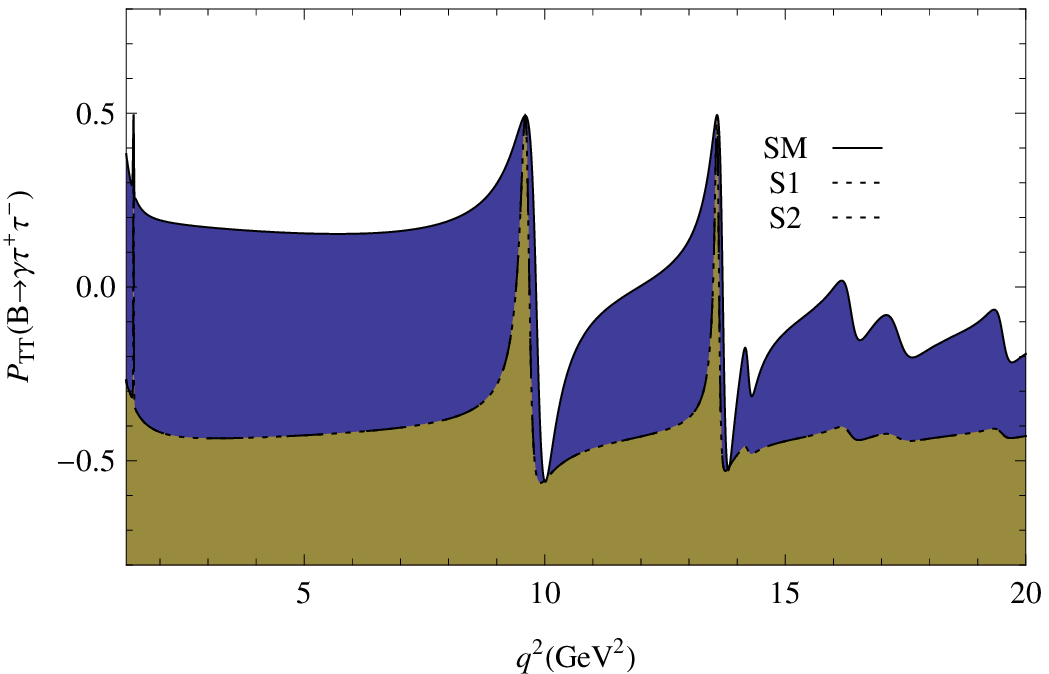}
\caption{ The dependence of the $P_{LL}$,$P_{NN}$ and $P_{TT}$
polarizations of the $B \rar\gamma l^+ l^- $ decay on
$q^2$ for $\mu$ and $\tau$ leptons.\label{fig2}}
\end{figure}

Finally, we shall present the expressions of the polarized forward-backward
asymmetry, which are very sensitive to the new physics effects. The explicit
expressions for ${\cal A}_{FB}$ are gives as:

\bea \label{e24} {\cal A}_{FB}^{LL} \es
\frac{1}{\Delta} \Bigg\{
- 4 m_B^2 \hat{s} (1-\hat{s})^2 v \mbox{\rm Re}
[A_1^\ast A_2 - B_1^\ast B_2] \nnb \\
\ar \frac{4}{\hat{m}_\ell v} f_B m_B \hat{s} (1-\hat{s}) (1-v^2)
\ln [1-v^2] \mbox{\rm Re}[(A_2^\ast-B_2^\ast) F]\Bigg\} ~, \nnb \\
\label{e6718} {\cal A}_{FB}^{LN} \es
\frac{1}{\Delta} \Bigg\{
- \frac{2}{3 \hat{m}_\ell} m_B^2 \sqrt{\hat{s}^3}(1-\hat{s})^2 v
(1-v^2) \Big( \mbox{\rm Im}[A_1^\ast B_1 + A_2^\ast B_2] \Big) \nnb \\
\ek f_B m_B^2 \sqrt{\hat{s}} (1-\hat{s}) \Big( + m_B \mbox{\rm
Im}[(A_1^\ast-A_2^\ast-B_1^\ast-B_2^\ast) F - \hat{s}
(A_1^\ast+A_2^\ast-B_1^\ast+B_2^\ast) F]\Big) I_7 \Bigg\}~, \nnb \\
\label{e6719} {\cal A}_{FB}^{NL} \es
\frac{1}{\Delta} \Bigg\{
+ \frac{2}{3 \hat{m}_\ell} m_B^2 \sqrt{\hat{s}^3}(1-\hat{s})^2 v
(1-v^2) \Big( -\mbox{\rm Im}[A_1^\ast B_1 + A_2^\ast B_2] \Big) \nnb \\
\ek m_B \mbox{\rm Im}[(A_1^\ast+A_2^\ast-B_1^\ast+B_2^\ast) F -
\hat{s} (A_1^\ast-A_2^\ast-B_1^\ast-B_2^\ast) F] I_7 \Big\}
 \Bigg\}~, \nnb \\
\label{e6720} {\cal A}_{FB}^{LT} \es
\frac{1}{\Delta} \Bigg\{
\frac{4}{3 \sqrt{\hat{s}}} \hat{m}_\ell (1-\hat{s})^2 \Big[ m_B^2
\hat{s} \ga \vel A_1 \ver^2 + \vel A_2 \ver^2 + \vel B_1 \ver^2 +
\vel B_2 \ver^2 \dr \Big] \nnb \\
\ar \frac{8}{3} m_B^2 \hat{m}_\ell \sqrt{\hat{s}} (1-\hat{s})^2
\Big( \mbox{\rm Re}[A_1^\ast B_1 + A_2^\ast B_2]\Big) \nnb \\
\ar \frac{1}{\sqrt{\hat{s}}} f_B^2 m_B^4 \hat{m}_\ell (1-\hat{s})
\Big[ (1-\hat{s}) \ga \vel F \ver^2 \dr
({\cal J}_{1} + {\cal J}_{2}) \Big]\nnb \\
\ek f_B m_B^3 \sqrt{\hat{s}} (1-\hat{s}^2) v^2 \mbox{\rm
Re}[(A_2^\ast-B_2^\ast) F^\ast]
{\cal J}_{4} \nnb \\
\ar f_B m_B^3 \sqrt{\hat{s}} (1-\hat{s})^2 (2-v^2) \mbox{\rm
Re}[(A_1^\ast+B_1^\ast) F] {\cal J}_{4} \Bigg\}~. \nnb \\
\eea

${\cal J}_i$ represent the following integrals
\bea {\cal J}_i \es
\int_{0}^{+1} {\cal G}_i(z) dz -
               \int_{-1}^{0} {\cal G}_i(z) dz ~,\nnb
\eea where \bea
\begin{array}{llll}
{\cal G}_{1} = \ds\frac{z \sqrt{1-z^2}}{(p_1 \cdot k) (p_2 \cdot
k)}~,& {\cal G}_{2} = \ds\frac{z \sqrt{1-z^2}}{(p_1 \cdot k)^2}~,&
{\cal G}_{3} = \ds\frac{\sqrt{1-z^2}}{(p_1 \cdot k)^2}~,& {\cal
G}_{4} = \ds\frac{z \sqrt{1-z^2}}{(p_1 \cdot k)}~.
\end{array} \nnb
\eea

\begin{figure}
\includegraphics[scale=0.8]{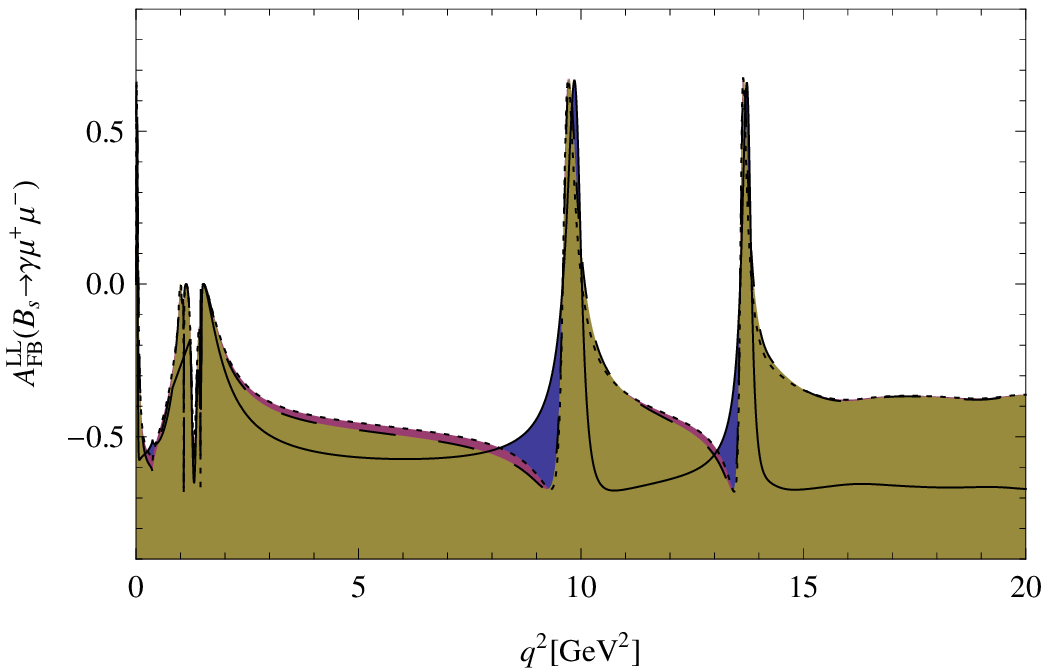}
\includegraphics[scale=0.8]{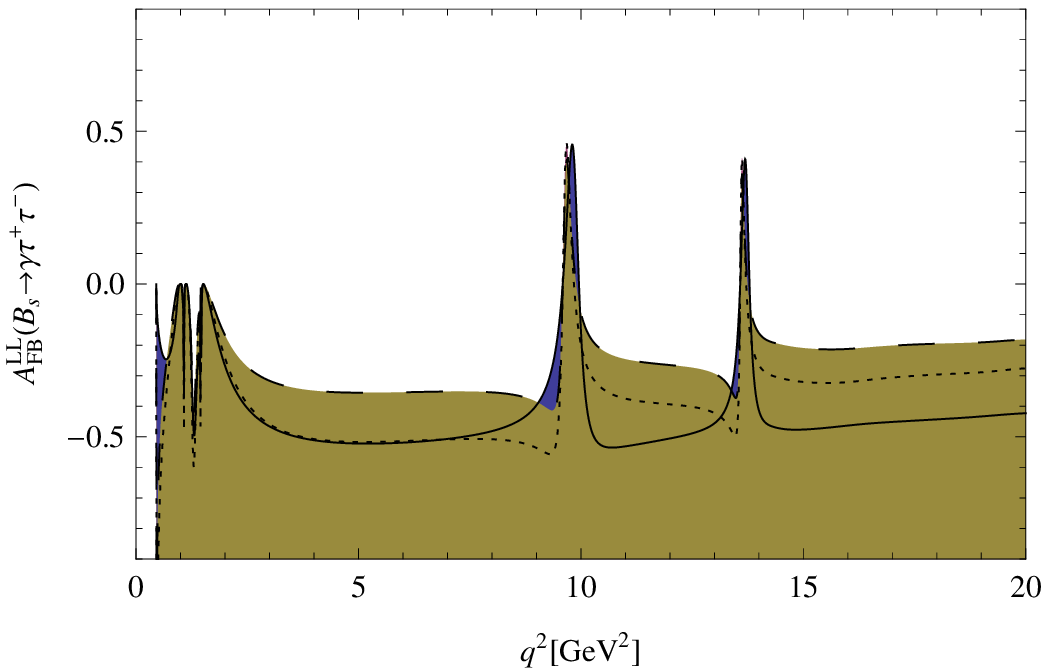}
\caption{ The dependence of the ${\cal A}_{FB}^{LL}$ polarizations
of the $B \rar \gamma l^+ l^- $ on $q^2$ for $\mu$ and
$\tau$ leptons.\label{fig3}}
\end{figure}

\section{Numerical Analysis}

For performing the numerical analysis, we use the following input
parameters entering into the expressions of the branching ratio,
double lepton polarization and forward-backward asymmetries:
$m_t=(173.5 \pm 0.6)~GeV$, $m_b=(4.8\pm0.1)~GeV$, $m_c=(1.46 \pm
0.05)~GeV$, $m_W = (80.385\pm 0.015)~GeV$, $m_\mu =
(105.658)\times 10^{-3}~GeV$, $\vel V_{tb} V_{ts}^\ast\ver =
0.041$, and $G_F=1.17 \times 10^{-5}~GeV^2$.

Among the remaining input parameters of the family non-universal
Z' model, the parameters $\vel B_{sb}^L \ver$, $B_{\ell\ell}^L$
and $B_{\ell\ell}^R$ are fixed from the results of UT
Collaboration \cite{R22}.

\begin{table}[h!]
\renewcommand{\arraystretch}{1.5}
\addtolength{\arraycolsep}{3pt}
$$
\begin{array}{|l|c|c|c|c|}
\hline \hline
             & \vel B_{sb}^L \ver \times 10^{-3} & \varphi_S^{L(0)} &
B_{\mu\mu}^L\times 10^{-2} & B_{\mu\mu}^R\times 10^{-2} \\
 \hline
   S1 & 1.09 \pm 0.22  & -72 \pm 7
& -4.75 \pm 2.44 & 1.97 \pm 2.24 \\
 S2 & 2.20 \pm 0.15 & -82 \pm 4
& -1.83 \pm 0.82 & 0.68 \pm 0.85 \\
\hline \hline
\end{array}
$$
\caption{The values of the $Z^\prime$ model parameters for two different
scenarios obtained by UT Collaborations.}
\renewcommand{\arraystretch}{1}
\addtolength{\arraycolsep}{-1.0pt}
\end{table}

In Fig. 1 we present the dependence of the differential decay rate
on $q^2$ for $B_s \to \ell^+ \ell^- \gamma$ for the $e$, $\mu$ and
$\tau$ channels. For completeness, the result for the SM is also
shown in the same figure. From this figure we see that the
differential branching ratios seem to be quite sensitive to the
existence of $Z^\prime$, but only at "low $q^2$" region.
Therefore, careful analysis of differential decay rate in the low
$q^2$ region can be useful for establishing the existence of
$Z^\prime$ boson.

In fig. 2 we depict the dependence of the double lepton
polarizations $P_{LL}$, $P_{NN}$ and $P_{TT}$ on $q^2$ for the
$\mu$ and $\tau$ channels. It follows from these figures that at
low and high $q^2$ regions $Z^\prime$ gives considerable
contributions to the $P_{NN}$ and $P_{TT}$ polarizations. In this
region, a careful study of these polarizations can be useful in
conforming the existence of $Z^\prime$.

The dependence of the polarized forward-backward asymmetry ${\cal
A}_{LL}$ on $q^2$ for the $\mu$ and $\tau$ channels is presented
in Fig. 3. From these figures, we observe that at low $q^2$
region, the values of ${\cal A}_{LL}$ are different for both
channels in the SM and $Z^\prime$ models, especially for the $S2$
scenario. This observation can serve as a useful tool in search of
the $Z^\prime$ boson.

\section{Conclusion}

In the present work, we studied the $B_s \to \ell^+ \ell^- \gamma$
decay in the family of non-universal $Z^\prime$ model. We
investigated the possible contributions of the $Z^\prime$ boson to
the branching ratio, double lepton polarizations, as well as,
polarized forward-backward asymmetries. We observe that, studying
these polarization effects in the low energy region $4m_\ell^2 \le
q^2 \le 8.0~GeV^2$ can give useful information in discriminating
the contributions of the the $Z^\prime$ boson. The aforementioned
measurable quantities are quite sensitive to the $Z^\prime$
contributions and discrepancies between prediction of the SM and
the family of non-universal $Z^\prime$ models can be indication
for the existence of the $Z^\prime$ boson. We hope that the
measurement of this channel can possibly be realized in the near
future at LHCb, since its branching ratio is of the same order as
that of the $B_s \to \ell^+ \ell^-$, which has already been
observed at LHCb. Checking then the predictions of the family of
non-universal $Z^\prime$ model, might be of help to confirm the
existence of the $Z^\prime$ boson.

\vspace{2cm}

\textbf{Acknowledgements}

The author would like to thank T.M.Aliev and N.K.Pak for
invaluable comments and useful discussions.
\newpage
\renewcommand{\topfraction}{.99}
\renewcommand{\bottomfraction}{.99}
\renewcommand{\textfraction}{.01}
\renewcommand{\floatpagefraction}{.99}
%
%
\newpage
\renewcommand{\topfraction}{.99}
\renewcommand{\bottomfraction}{.99}
\renewcommand{\textfraction}{.01}
\renewcommand{\floatpagefraction}{.99}

\vspace{2cm}
\clearpage


\newpage
\begin{thebibliography}{100}

\bibitem{R01} R. Aaij et al.
  LHCb Colloboration Phys. rev. Lett. {\bf 110}, 021801 (2013).
%
\bibitem{R02} S. Chatrchyan {\it et al.}, CMS Collaboration,
  Phys. Rev. Lett. {\bf 111}, 101804 (2013).
%
\bibitem{R03} T. M. Aliev, A. Ozpineci, M. Savci,
  Phys. Rev. D {\bf 55}, 7059 (1997).
%
\bibitem{R04} E. Nardi,
  Phys. Rev. D {\bf 48}, 1240 (1993);
              V. Barger, M. S. Berger, and R. J. Phillips,
  Phys. Rev. D {\bf 52}, 1663 (1995).
%
\bibitem{R05} G.Buchalla, G. Burdman, C. T. Hill, and D. Comins,
  Phys. Rev. D {\bf 53}, 5185 (1996).
%
\bibitem{R06} P. Langacker and M. Plumacher,
  Phys. Rev. D {\bf 62}, 01013006 (2000).
%
\bibitem{R07} Q. Chang, X. Q. Li, Y. D. Yang,
  JHEP {\bf 082}, 1002 (2010).
%
\bibitem{R08} Q. Chang, X. Q. Li, Y. D. Yang,
  JHEP {\bf 1004}, 052 (2010).
%
\bibitem{R09} Q. Chang,
  Nucl. Phys. B {\bf 845}, 179 (2011).
%
\bibitem{R10} Y. Li, J. Hua, K. C. Yang,
  Eur. Phys. J. C {\bf 71}, 1775 (2011).
%
\bibitem{R11} T. M. Aliev, M. Savci,
  Nucl. Phys. B {\bf 863}, 398 (2012).
%
\bibitem{R12} N. Katirci, K. Azizi,
  J. Phys. G {\bf 40}, 085005 (2013).
%
\bibitem{R13} G.Buchalla, A.J. Buras, and M. E. Lautenbacher,
  Rev. Mod. Phys.{\bf 68}, 1125 (1996).
%
\bibitem{R14} M. Misiak,
  Nucl. Phys. B {\bf 393}, 23 (1993);
  Erratum {\it ibid.} B {\bf 439}, 461 (1995).
%
\bibitem{R15} C. H. Chen and H. Hatanaka,
  Phys. Rev. D {\bf 73}, 075003 (2006).
%
\bibitem{R16} V. Barger {\it et al.},
  Phys. Rev. D {\bf 80}, 055008 (2009);
              T. M. Aliev, M. Savci,
  Phys. Lett. B {\bf 718}, 566 (2012).
%
\bibitem{R17} A. Buras and M. M\"{u}nz,
  Phys. Rev. D {\bf 52}, 186 (1995).
%
\bibitem{R18} C.H. Chen, C.Q. Geng,
  Phys. Rev. D {\bf 63} 054005 (2001).
%
\bibitem{R19} T. M. Aliev, N.K. Pak and M. Savci,
  Phys. Lett. B {\bf 424}, 175 (1998)
%
\bibitem{R20} U. O. Yilmaz, B. B. Sirvanli, and G. Turan,
  Nucl. Phys. B {\bf 692}, 249 (2004).
%
\bibitem{R21} K. Azizi, N.K. Pak, B. B. Sirvanli,
  JHEP, {\bf 034}, 1202 (2012).
%
\bibitem{R22} M.Bona {\it et al.},
  UT-fit Collaboration, PMC Phys. A{\bf 3}, 6 (2009).
%
\end{thebibliography}
\end{document}